\documentclass[aip,jcp]{revtex4-1}

\usepackage{amsmath}
\usepackage{amssymb}
\usepackage{graphicx}
\usepackage{soul}

\def\bea{\begin{eqnarray}}
\def\eea{\end{eqnarray}}
\def\ben{\begin{equation}}
\def\een{\end{equation}}
\def\benu{\begin{enumerate}}
\def\enu{\end{enumerate}}

\def\n{n}
\def\lsim {\ifmmode {\buildrel<\over\sim}}

\def\sss{\scriptscriptstyle\rm}

\def\1var{(\bx_1...\bx\N)}

\def\half{\frac{1}{2}}

\def\br{{\bf r}}

\def\b1{{\bf 1}}
\def\bx{{x}}

\def\x{_{\sss X}}
\def\c{_{\sss C}}

\def\xc{_{\sss XC}}
\def\Hx{_{\sss HX}}
\def\Hxc{_{\sss HXC}}

\def\N{_{\sss N}}
\def\H{_{\sss H}}

\def\sph_int{ {\int d^3 r}}

\def\infintd3r{ \int_{-\infty}^\infty d^3r\,}
\def\intd3r{ \int d^3r\,}

\def\laplace1d{\frac{d^2}{dx^2}}
\def\plaplace1d{\frac{d^2}{d{x'}^2}}

\def\padr2{\frac{\partial^2}{\partial r^2}}

\def\N{{\cal N}}

\def\b{{\beta}}

\begin{document}


\title{Density Functional Resonance Theory: complex density functions, convergence, orbital energies, and functionals}


\author{Daniel L. Whitenack}
\email[]{dwhitena@purdue.edu}
\homepage[]{http://www.purdue.edu/dft}
\affiliation{Department of Physics, Purdue University, 525 Northwestern Avenue, West Lafayette, IN 47907, USA}
\author{Adam Wasserman}
\email[]{awasser@purdue.edu}
\affiliation{Department of Chemistry, Purdue University, 560 Oval Drive, West Lafayette, IN 47907, USA}
\affiliation{Department of Physics, Purdue University, 525 Northwestern Avenue, West Lafayette, IN 47907, USA}


\date{\today}

\begin{abstract}
Aspects of Density Functional Resonance Theory (DFRT) [Phys. Rev. Lett. \textbf{107}, 163002 (2011)], a recently developed complex-scaled version of ground-state Density Functional Theory (DFT), are studied in detail.  The asymptotic behavior of the complex density function is related to the complex resonance energy and system's threshold energy, and the function's local oscillatory behavior is connected with preferential directions of electron decay.  Practical considerations for implementation of the theory are addressed including sensitivity to the complex-scaling parameter, $\theta$.  In Kohn-Sham DFRT, it is shown that almost all $\theta$-dependence in the calculated energies and lifetimes can be extinguished via use of a proper basis set or fine grid.   The highest occupied Kohn-Sham orbital energy and lifetime are related to a physical affinity and width, and the threshold energy of the Kohn-Sham system is shown to be equal to the threshold energy of the interacting system shifted by a well-defined functional.  Finally, various complex-scaling conditions are derived which relate the functionals of ground-state DFT to those of DFRT via proper scaling factors and a non-Hermitian coupling constant system.  
\end{abstract}


\maketitle


\section{Introduction}
Density Functional Theory (DFT) \cite{hohenberg, kohn, parr} has provided one of the most accurate and reliable methods to calculate the electronic properties of molecules, clusters, and materials from first principles. It is the most widely used formalism in modern computational quantum chemistry \cite{Martin}. In addition, DFT's time-dependent extension (TDDFT) \cite{RG84} can now be applied to a wealth of excited-state and time-dependent properties in both linear and non-linear regimes \cite{tddftnew}. An advantage of DFT over wavefunction-based methods is that the one-body electron density is the primary variable, not the intracable many-body wavefunction.  

Recently, a complex-scaled version of DFT, Density Functional Resonance Theory (DFRT), was introduced~\cite{WW11} that allows one to calculate the in-principle exact resonance energy and lifetime of the Lowest Energy Resonance (LER) of an unbound system.  This theory makes use of a complex ``Kohn-Sham'' system that facilitates self-consistent calculations on many-electron systems with a complex-valued density function as the primary variable.  As shown in Refs.~\cite{WW10,WW11} and in this work, the use of a localized complex density is advantageous, especially in the case of resonances where the wavefunction is not only an awkward many-body object but also divergent, or non-normalizeable, because of the transient nature of the system.  As we will argue here, DFRT is a promising approach to calculate negative electron affinities (NEA) and resonance energies and lifetimes.

Shape and Feshbach resonances in low-energy electron scattering processes \cite{palmer,S08,*S11,FGMS07} are of growing interest in biological systems \cite{PCHS03,PS05,MBBF10}, atmospheric sciences, lasers, and astrophysics \cite{massey,LSBL02,S10,OSAW11}.  For example, consider electron impact on DNA.  Experimentalists have made great strides in their understanding of damage to DNA caused by secondary low-energy electrons during cancer radiation.  It turns out that this secondary process is much more important than previously thought~\cite{S09}.  Next to these experimental efforts, theoretical calculations are either very limited or extremely non-trivial due to the size of the relevant species, the many-body nature of the problem, and the bound-free correlations.  Even in small systems such as the molecular hydrogen anion or doubly negative ions, stability and structure are still a very active subject of research~\cite{S11, JKGS11}.  

In addition to relevant metastable properties that are within reach of the theory, DFRT reduces to standard ground-state DFT with the removal of the complex-scaling transformation and thus allows one to examine current topics of research in approximate DFT from a different viewpoint.  One example of such an application is recent work investigating derivative discontinuities of the energy at the maximum number of bound electrons~\cite{WZW11}.

In this paper, several aspects of the DFRT formalism, introduced in Ref.~\citenum{WW11}, are studied in detail.  The outline of the paper is as follows.  First, in section~\ref{sec:dfrt} we review the analog of Hohenberg and Kohn's theorems for the LER and the main ideas of DFRT.  Aspects of convergence in DFRT calculations are addressed in section~\ref{sec:prac}.  Section~\ref{sec:ntheta} provides a discussion of the physical significance and use of the complex density function, the primary variable of DFRT. The Kohn-Sham orbital energies and lifetimes are related to physical properties of the interacting system in section~\ref{sec:orb}, and possible approximations for the complex universal energy functional are explored in section~\ref{sec:func}.  Lastly, we conclude and mention some promising applications and extensions of DFRT.     

\section{\label{sec:dfrt} Density Functional Resonance Theory}
To begin, we give a brief sketch of the analogs of the Hohenberg-Kohn (HK) theorems~\cite{WM07} for a LER, and we describe the main ideas of Kohn-Sham DFRT.  

A quantum resonance is associated with a complex number, $E_n = {\cal{E}}_n - (\Gamma_n /2)i$, that is a pole of the scattering matrix or a peak in the continuum density of states. The index $n$ labels a specific resonance (atomic units are used throughout).  ${\cal{E}}_n$ is the resonance energy or position, and $\Gamma_n$ is the resonance width.  The lifetimes of the resonances, ${\cal{L}}_n$, are given by the inverse of the widths.  Consider the LER of an unbound system and assume that this lowest energy metastable state is also the longest-lived resonance (this condition is not proven, but is known to be the typical case~\cite{S73a,S73b}).  

The complex ``density'' associated with the LER is
\ben
n_\theta(\br)=\langle \Psi_\theta^L|\hat{n}(\br)|\Psi_\theta^R\rangle
\label{eq:cden}
\een
where $\hat{n}(\br)$ is the density operator, and $\langle\Psi_\theta^L|$ and $|\Psi_\theta^R\rangle$ are the left and right eigenvectors of the complex scaled Hamiltonian, $\hat{H}_\theta$, corresponding to the LER. The angle $\theta$ is the complex scaling parameter in the transformation of the coordinates from $\textbf{r}$ to $\textbf{r} e^{i \theta}$ (see Ref.~\citenum{M11} for a review of such transformations).

In general, there is no variational principle for the complex energy associated with the LER, but only a \emph{stability} principle.  In Ref.~\citenum{WM07}, a complex density variational principle was presented for the LER pertaining to trial functions of a certain kind.  Specifically those trial functions, $\Phi_{trial}$, which can be expanded in a set of resonance wavefunctions and for which $|\langle \Psi_\theta^L |\Phi_{trial}^R\rangle|>1/2$.  In addition, Ernzerhof and co-workers have shown that for potentials and densities with a small imaginary part (compared to their real part), the standard density variational principle is maintained~\cite{E06}.    

A one-to-one correspondence between complex densities and complex-scaled potentials can be shown via contradiction, but the Levy-Lieb constrained search approach is taken here.  In this approach, the lowest-energy resonance energy and lifetime of a system defined by the Hamiltonian, $\hat{H}=\hat{T}+\hat{V}_{ee}+\int d\br \hat{n}(\br) v(\br)$, can be written as:
\begin{eqnarray}
 E_{\sss LER} &=& \underset{\Psi_\theta}{\mbox{min}} \left( \begin{array}{c} \mbox{Re} \\ -2 \mbox{Im} \end{array} \right) \nonumber \\ & & \times \langle \Psi_\theta^L|\hat{T^\theta}+\hat{V}_{ee}^\theta+\int d\br \hat{n}(\br) v(\br e^{i \theta}) |\Psi_\theta^R\rangle
\end{eqnarray}
where the energy and width are minimized over all resonance wavefunctions.  Next, split the minimization into two steps.  First, search over all resonance wavefunctions that give a certain complex density, and then search over all complex densities.
\begin{widetext}
\begin{eqnarray}
 E_{\sss LER} &=& \underset{n_\theta}{\mbox{min}} \left[ \underset{\Psi_\theta \rightarrow n_\theta}{\mbox{min}} \left( \begin{array}{c} \mbox{Re} \\ -2 \mbox{Im} \end{array} \right) \langle \Psi_\theta^L|\hat{T^\theta}+\hat{V}_{ee}^\theta+\int d\br \hat{n}(\br) v(\br e^{i \theta}) |\Psi_\theta^R\rangle \right] \nonumber \\
 &=& \underset{n_\theta}{\mbox{min}} \left( \begin{array}{c} \mbox{Re} \\ -2 \mbox{Im} \end{array} \right) \left( F^\theta[n_\theta] +\int d\br \ n_\theta(\br) v(\br e^{i \theta}) \right) 
\label{eq:constsearch}
\end{eqnarray}
\end{widetext}
Here, $F^\theta[n_\theta]$ is a complex-valued ``universal functional'' defined by,
\begin{equation}
 F^\theta[n_\theta] = \underset{\Psi_\theta \rightarrow n_\theta}{\mbox{min}} \left( \begin{array}{c} \mbox{Re} \\ -2 \mbox{Im} \end{array} \right) \langle \Psi_\theta^L|\hat{T^\theta}+\hat{V}_{ee}^\theta |\Psi_\theta^R \rangle
 \label{eq:fn}
\end{equation}
Carrying out the minimization with respect to variations in the density given the constraint that the density integrates to the number of electrons ($\mu$ - Lagrange multiplier) gives
\begin{equation}
 \frac{\delta}{\delta n_\theta} \left[  F^\theta[n_\theta] + \int d\br \ n_\theta(\br) v(\br e^{i \theta}) - \mu \int d\br \ n_\theta (\br) \right] = 0
\end{equation}
and one obtains the result,
\begin{equation}
 v(\br e^{i \theta}) = \mu - \frac{\delta F^\theta[n_\theta]}{\delta n_\theta} 
\end{equation}
Therefore, $v(\br e^{i \theta})$ maps directly to the specific $n_\theta (\br)$ that satisfies the above condition.

In the original extension of the Hohenberg-Kohn Theorems to LER's~\cite{WM07}, both a proof by contradiction and the Levy-Lieb type constained search were employed.  In the proof by contradiction, it is assumed that the energy and width are minimized over all trial complex densities that are $v^{\theta}$-representable.  This means that the trial densities can be associated with a Hamiltonian containing the smooth complex-scaled external potential $v(\br e^{i \theta})$.  However, if an arbitrary trial complex density is chosen, it is not immediately evident that this condition is fulfilled.  The Levy-Lieb constrained search of Eq.~\ref{eq:constsearch} lifts this  restriction and replaces it with a weaker restriction on the trial complex densities.  The search in Eq.~\ref{eq:constsearch} is performed over all $N$-representable densities.  This is very similar to the $N$-representability condition normally applied in ground-state DFT (see Ref.~\citenum{gross1995density} for a review of $v$-representability and $N$-representability in ground-state DFT), but here we search over all localized \emph{complex-valued} functions that integrate to the number of particles, rather than searching over purely real functions.  This is a larger space of densities than what is taken in ground-state DFT and hence the $N$-representability condition of DFRT is a weaker restriction than that of $N$-representability in DFT.

Having motivated this correspondence between complex densities and complex-scaled potentials, we proceed with defining the theory.  It is required that $n_\theta(\br)$, corresponding to the LER, be normalized to the number of electrons, as real densities are:
\ben
\int d\br \ n_\theta(\br)=N
\label{eq:norm}
\een
The energy and lifetime of the resonance can be extracted from $n_\theta$ with a properly scaled energy functional.  For $N$ electrons we write this functional as
\begin{eqnarray}
{\cal{E}}[n_\theta]-\frac{i}{2}{\cal{L}}^{-1}[n_\theta] &=& T_s^{\theta}[n_\theta]+\int d\br n_\theta(\br) v(\br e^{i\theta}) \nonumber \\
 & & + E\H^{\theta}[n_\theta] + E\xc^{\theta}[n_\theta] 
\label{e:chypothesis}
\end{eqnarray}
where $v$ is the external potential, $T_s^{\theta}$ the complex-scaled non-interacting Kinetic energy functional, $E\H^\theta$ is the classical Hartree energy functional, and $E\xc^\theta$ is the exchange-correlation energy functional.  We require that $T_s^\theta[n_\theta]=e^{-2i\theta} T_s[n_\theta]$  and $E\H^\theta[n_\theta]=e^{-i\theta}E\H[n_\theta]$, where $T_s[n_\theta]$ and $E\H[n_\theta]$ are the standard non-interacting kinetic energy and Hartree functionals evaluated at the complex densities.  Without an explicit expression for $E\xc^\theta[n_\theta]$, however, the total energy cannot be calculated via Eq.~\ref{e:chypothesis}.  The exact scaling of functionals and possible candidates for approximate functionals will be discussed in Section~\ref{sec:func}.

Now map the system of interacting electrons whose LER density is $n_\theta(\br)$ to one of $N$ particles 
moving independently in a complex ``Kohn-Sham'' potential $v_s^\theta(\br)$ defined such that its $N$ occupied orbitals 
$\{\phi_i^\theta(\br)\}$ yield the interacting LER-density via 
$n_\theta(\br)=\sum_{i=1}^N \langle\phi_i^{\theta,L}|\hat{n}(\br)|\phi_i^{\theta,R}\rangle$. In Moiseyev's Hermitian representation of complex-scaling~\cite{M83}, the complex Kohn-Sham equations are:
\ben
\left(\begin{array}{cc}
\hat{h}_1 - \varepsilon_i & -\hat{h}_2 - 2\tau_i^{-1}\\
\hat{h}_2 + 2\tau_i^{-1} & \hat{h}_1 - \varepsilon_i\\
      \end{array}\right)
\left(\begin{array}{c}
{\mbox{Re}(\phi_i^{R,\theta})}\\
{\mbox{Im}(\phi_i^{R,\theta})}\\
      \end{array}\right)
= 0
~~,
\label{eq:cks}
\een
where $\hat{h}_1 = -\half\cos(2\theta)\nabla^2 + \mbox{Re}(v_s^\theta(\br))$, and $\hat{h}_2=\half\sin(2\theta)\nabla^2+ \mbox{Im}(v_s^\theta(\br))$. The set of $\{\varepsilon_i\}$ and $\{\tau_i\}$ are the orbital resonance energies and lifetimes of the
Kohn-Sham particles, and the set of $\{\phi_i^{R,\theta}\}$ and $\{\phi_i^{L,\theta}\}$ are the right and left eigenfunctions of the non-Hermitian Kohn-Sham Hamiltonian.

To show that the solution of Eq.~\ref{e:chypothesis} can be written in an Euler-Lagrange form, expand the right and left Kohn-Sham orbitals in an orthonormal basis,
\begin{equation}
 \phi_i^{R,\theta} (\br) = \displaystyle\sum\limits_{i}^N c_i^R \chi_i(\br) \ ; \quad \phi_i^{L,\theta}(\br) = \displaystyle\sum\limits_{i}^N c_i^L \chi_i(\br)
\end{equation}
The complex density can then be written as,
\begin{equation}
 n_\theta(\br)= \displaystyle\sum\limits_{i}^N c_i^L c_i^R \chi_i^2(\br)
\end{equation}
and the functional derivative of $E^{\theta}[n_\theta]={\cal{E}}[n_\theta]-\frac{i}{2}{\cal{L}}^{-1}[n_\theta]$ with respect to $n_\theta(\br)$ has the form
\begin{eqnarray}
 \frac{\delta E^\theta[n_\theta]}{\delta n_\theta(\br)} &=& \frac{1}{2 c_1^L \chi_1^2(\br)} \frac{\partial E^\theta}{\partial c_1^R} + \frac{1}{2 c_1^R \chi_1^2(\br)} \frac{\partial E^\theta}{\partial c_1^L} \nonumber \\
 & & + \frac{1}{2 c_2^L \chi_2^2(\br)} \frac{\partial E^\theta}{\partial c_2^R} + \ldots
\end{eqnarray}
Moiseyev \emph{et al.} have shown that the resonance wavefunction is given by setting $\partial E^\theta / \partial c_i^R = \partial E^\theta / \partial c_i^L = 0$ for all $i$~\cite{M11}.  Therefore, the LER solution of Eq.~\ref{e:chypothesis}, with the constraint that the complex density integrates to the number of particles, is given by,
\ben
\frac{\delta E_\theta[n_\theta]}{\delta n_\theta}-\mu\int d\br n_\theta(\br) = 0~~.
\een 
Performing the variarion in Eq.~\ref{e:chypothesis} and comparing with Eq.~\ref{eq:cks} leads to an expression for the Kohn-Sham potential that is again analogous to that of standard KS-DFT:
\ben
v_s^\theta(\br)=v(\br e^{i\theta})+e^{-i\theta}v\H[n_\theta](\br)+v\xc^\theta[n_\theta](\br)~~,
\label{eqn:vs}
\een
where $v\xc^\theta[n_\theta](\br)=\delta E\xc^\theta[n_\theta]/\delta n_\theta(\br)|_{\rm LER}$.  Now with this auxiliary Kohn-Sham system established, self-consistent calculations for the LER's energy and lifetime are possible with an intial guess for the complex density.

\section{\label{sec:prac} Aspects of Convergence in DFRT Calculations}

The main structure of the DFRT formalism mimics that of ground-state DFT.  One chooses an initial guess for the complex density, calculates the KS potential from Eq.~\ref{eqn:vs}, solves the KS equations of Eq.~\ref{eq:cks}, constructs a new density by summing the KS orbitals, and repeats until self-consistency is reached.  Despite the similarities, Kohn-Sham DFRT is a complex-scaled Non-Hermitian formalism, and a number of unique practical aspects of the theory should be discussed.  

The resonance energy and lifetime is independent of $\theta$ in complex-scaling theory~\cite{moiseyev1}.  However when doing a numerical calculation, the use of finite basis sets or finite grids necessitates finding an optimum $\theta$.  This optimum condition is acheived by examining $\theta$-trajectories~\cite{moiseyev1}.  When the real and imaginary parts of the resonance energy become stationary around the optimum value of $\theta$ one can see a kink or loop in the trajectory. With many more grid points or a larger basis this procedure becomes less important as all the $\theta$-trajectory points collapse to the region of the resonance.

\begin{figure} 
\scalebox{0.45}{\includegraphics{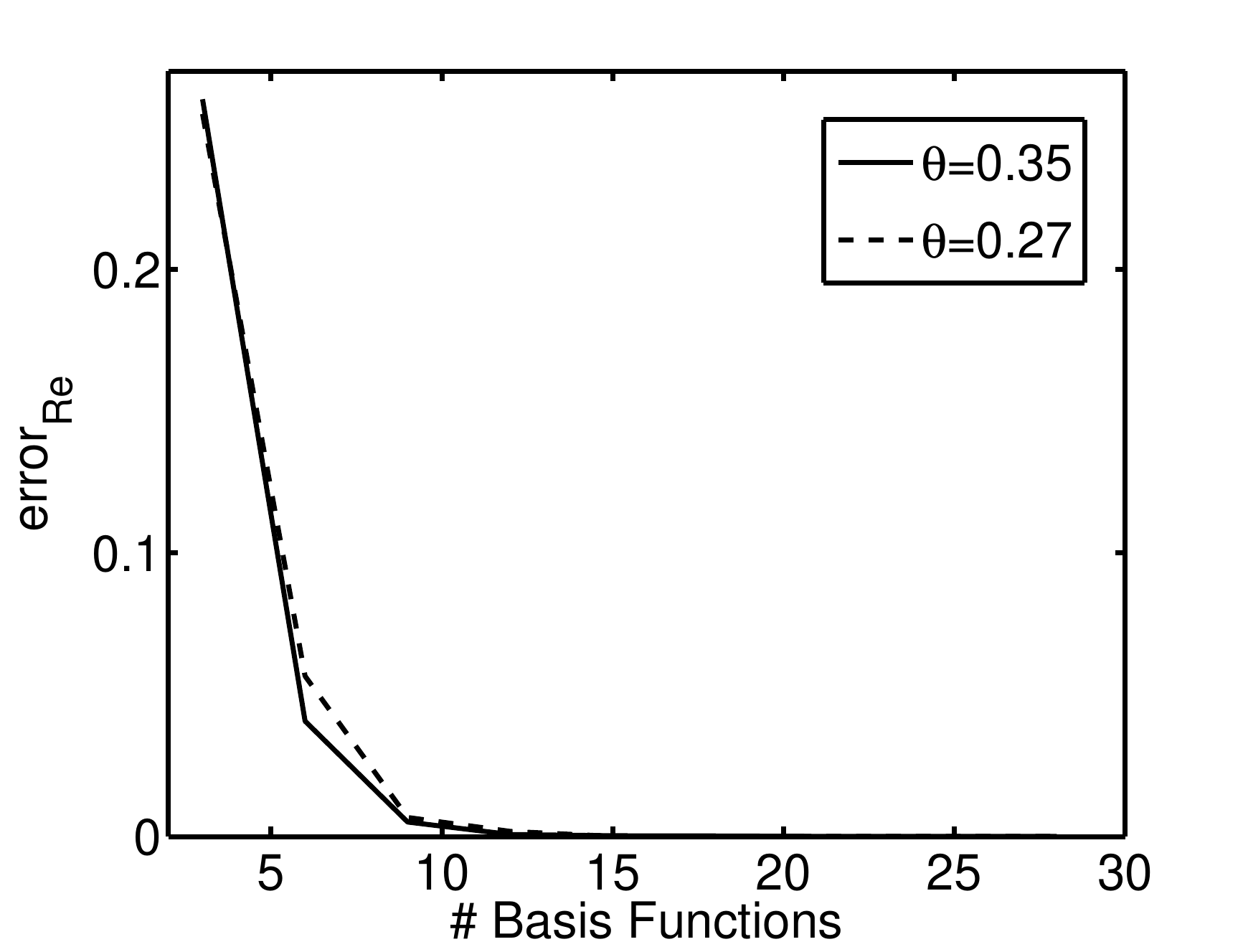}} 
\scalebox{0.45}{\includegraphics{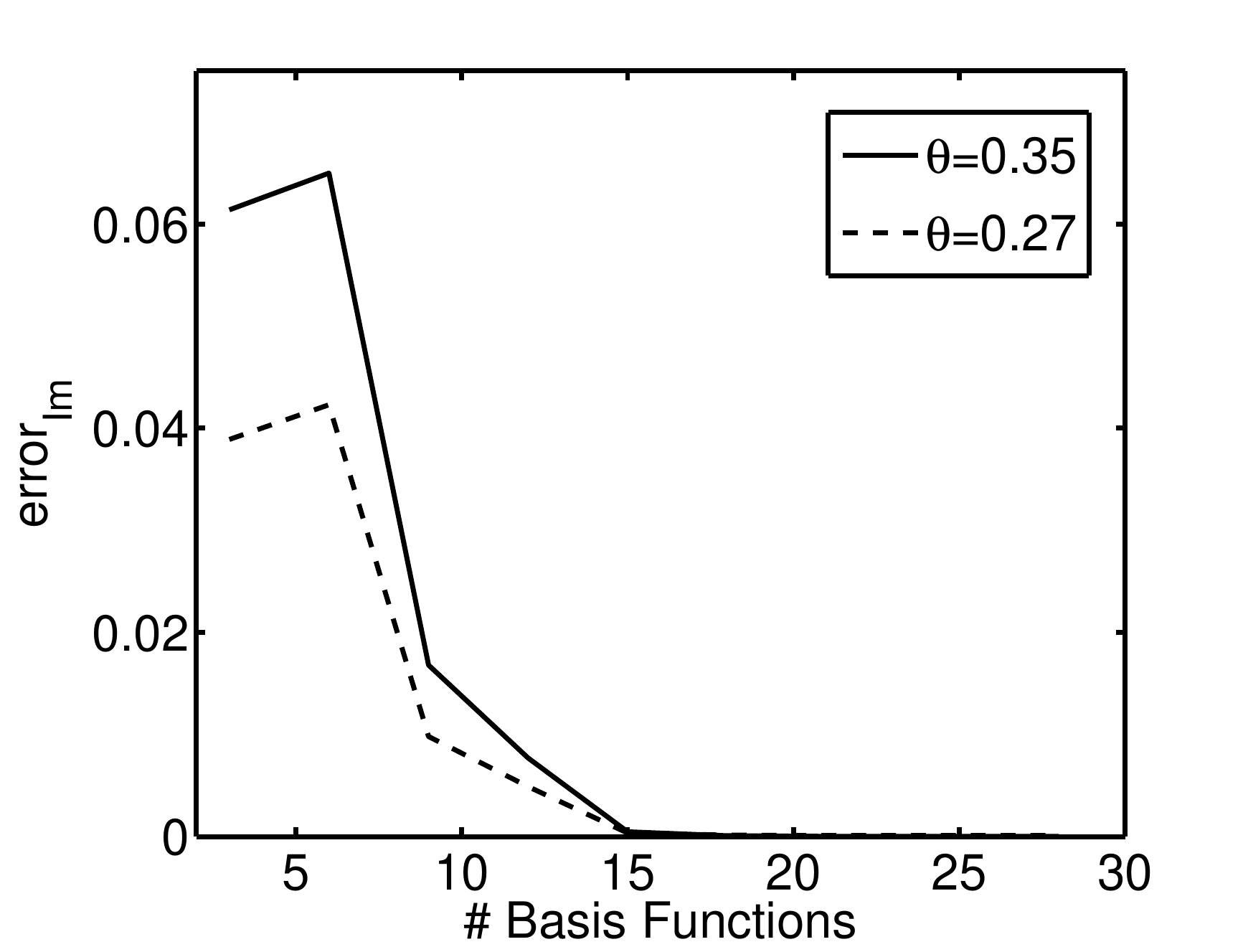}}
\caption{\label{fig:pib}The difference between the real and imaginary parts of $n_\theta$ from the basis function and the fourier grid method.  Given a choice of $\theta$ these converge to the same density. ($a=4$, $b=0.5$, $c=4$, and $d=2$)}
\end{figure}

$\theta$-independence of the energy is preserved by the SCF procedure of DFRT. As the grid or basis size increases the dependence on $\theta$ becomes negligible.  To illustrate this $\theta$-independence, calculations were performed on a model potential similar to the one discussed in Ref.~\citenum{WW11} filled with two solf-coulomb interacting electrons.  The potential has the form, 
\begin{equation}
v(x) = a\left[\sum\limits^{2}_{j=1} \left(1+e^{-2 c (x + (-1)^jd)}\right)^{-1} \right]-\alpha e^{-\frac{x^2}{b}}
\label{eq:modelpot}
\end{equation} 
where $a$, $\alpha$, $b$, $c$, and $d$ are constants.  Results for the energy and lifetime of a two-electron resonance were obtained using the Fourier Grid Hamiltonian (FGH) method and with both a particle-in-a-box and harmonic oscillator basis.  It can be seen that the method of using $n_\theta$ in a scaled functional is independent of $\theta$, as long as the basis is large enough.  In other words, for a given choice of $\theta$ the density obtained using a large grid and a large basis will be the same, and those densities will yield the same result in the scaled functional. The error in the density, plotted in Fig.~\ref{fig:pib} versus the number of basis functions, was calculated as,
\begin{eqnarray}
 \mbox{error}_{\sss Re} &=& \int dx \ ( \mbox{Re}(\n_\theta^{pib}(x)) -\mbox{Re}(\n_\theta^{fgh} (x)) )^2  \nonumber \\
 \mbox{error}_{\sss Im} &=& \int dx \ ( \mbox{Im}(\n_\theta^{pib}(x)) -\mbox{Im}(\n_\theta^{fgh} (x)) )^2
\end{eqnarray}
 where $\n_\theta^{fgh}$ is an accurate density from the Fourier grid Hamiltonian method and $\n_\theta^{pib}$ is the density obtained using the particle-in-a-box basis (both obtained with the same $\theta$).  Figure~\ref{fig:pib} shows the convergence of the density as the number of basis functions increases.  
Note that if one chose a smaller basis or grid, an optimum $\theta$ could still be found for the calculation.  This optimum $\theta$ gives the correct resonance energy and lifetime~\cite{moiseyev1,reinhardt,simon,balslev}.  This principle of $\theta$-independence is important, because within a Kohn-Sham DFRT calculation one has to make a choice for the scaling parameter. Yet, the advantage of the scheme is that one only has to solve one-body, non-interacting equations.  For such equations, one should be able to efficiently use a large enough basis set or a fine enough grid to extinguish most of the numerical $\theta$-dependence. Thus, a well-known drawback of the complex-scaling technique \cite{moiseyev1,reinhardt,simon} is outdone by the benefit of never having to deal with $N$-particle wavefunctions, but just one-body (complex) densities.

Although the energy in DFRT is theoretically independent of the scaling parameter, all of the potentials and the density itself depend on the choice of $\theta$.  Therefore, one should be careful to compare only potentials and densities calculated with the same $\theta$. 

In addition to the issue of $\theta$-dependence, convergence of the solutions should be discussed.  As can be seen in Fig.~\ref{fig:pib} and in Ref.~\citenum{WW10}, often the imaginary part of the energy has a slower convergence than the real part.  This occurs when the width ($\mbox{Im}(E^\theta)$) of the resonance is much smaller than the position ($\mbox{Re}(E^\theta)$) which is typically true for long-lived resonances.  One must evaluate the functional using a very accurate complex density in order to precisely calculate a narrow width.  However, the self-consistent DFRT method has already been tested for small model systems of interacting electrons where this accuracy is possible~\cite{WW10,WW11}.

\section{\label{sec:ntheta} The Complex Density Function}

A natural question is: what kind of physical significance can be associated with the real and imaginary parts of the complex density function?  This question has already been partially explored by Moiseyev and Barkay.  They showed that the phase of the square root of the complex density in systems like quantum dots or tunneling diodes is related to the phase of a measureable quantity, the complex tunneling probability amplitude~\cite{BM01}.  This quantity provides the probability for obtaining a specific scattering result such as transmission through a diode.  Using this property of the complex density, Moiseyev and Barkay derived a formula for the tunneling probability amplitude which correctly predicted the energetic position of resonances.  In addition to the study by Moiseyev and Barkay, Buchleitner \emph{et al} have examined how a properly normalized magnitude of the complex density can be interpreted as the real electron density of the metastable system~\cite{buchleitner, M11}.  The fact that the real part of the density might resemble a bound density for very long-lived states has been emphasized in previous work~\cite{WW10}.

In addition to these interpretations, the asymptotics of the LER's complex density contain relevant information about the real system.  As $r \rightarrow \infty$ the density, $n_\theta$, behaves like~\cite{M11}
\begin{equation}
 n_{\theta}(\textbf{r}) \sim A e^{i 2 \sqrt{2 \Delta E} e^{i \theta} r}
 \label{eq:nassym}
\end{equation}
Where $\Delta E = E^\theta - E^{th}$, $E^\theta$ is the complex energy of the LER, and $E^{th}$ is the threshold energy for the relevant decay channel.  The assymptotic expression for $n_\theta$ can be split up into real and imaginary parts.
\begin{eqnarray}
 \left( \begin{array}{c} \mbox{Re} \\ \mbox{Im} \end{array} \right) n_\theta (\br) &\sim& A e^{-2 \sqrt{2} \mbox{Im}(e^{i \theta} \sqrt{\Delta E}) r} \nonumber \\
& & \times \left( \begin{array}{c} \cos{2 \sqrt{2} \mbox{Re}(e^{i \theta} \sqrt{\Delta E}) r} \\ \sin{2 \sqrt{2} \mbox{Re}(e^{i \theta} \sqrt{\Delta E}) r} \end{array} \right)
\end{eqnarray}
One can see that the decay of the real part of $n_\theta$ is coupled to both the real and imaginary parts of $\Delta E$ as is the imaginary part of $n_\theta$.  If the density is written in the following way
\begin{equation}
 n_\theta (\br) = |n_\theta (\br)| e^{i \phi (\br)}
\end{equation}
where at large $r$,
\begin{equation}
 |n_\theta(\br)| \sim A e^{-2 \sqrt{2} \mbox{Im}(e^{i \theta} \sqrt{\Delta E}) r}
\end{equation}
\begin{equation}
 \phi(\br) \sim 2 \sqrt{2} \mbox{Re}(e^{i \theta} \sqrt{\Delta E}) r
\end{equation}
then the decay of the magnitude of $n_\theta$ is governed by only the imaginary part of $e^{i \theta} \Delta E$ and the decay of the phase of $n_\theta$ is governed by only the real part of $e^{i \theta} \Delta E$.   

\begin{figure}
\begin{center}
\scalebox{0.45}{\includegraphics{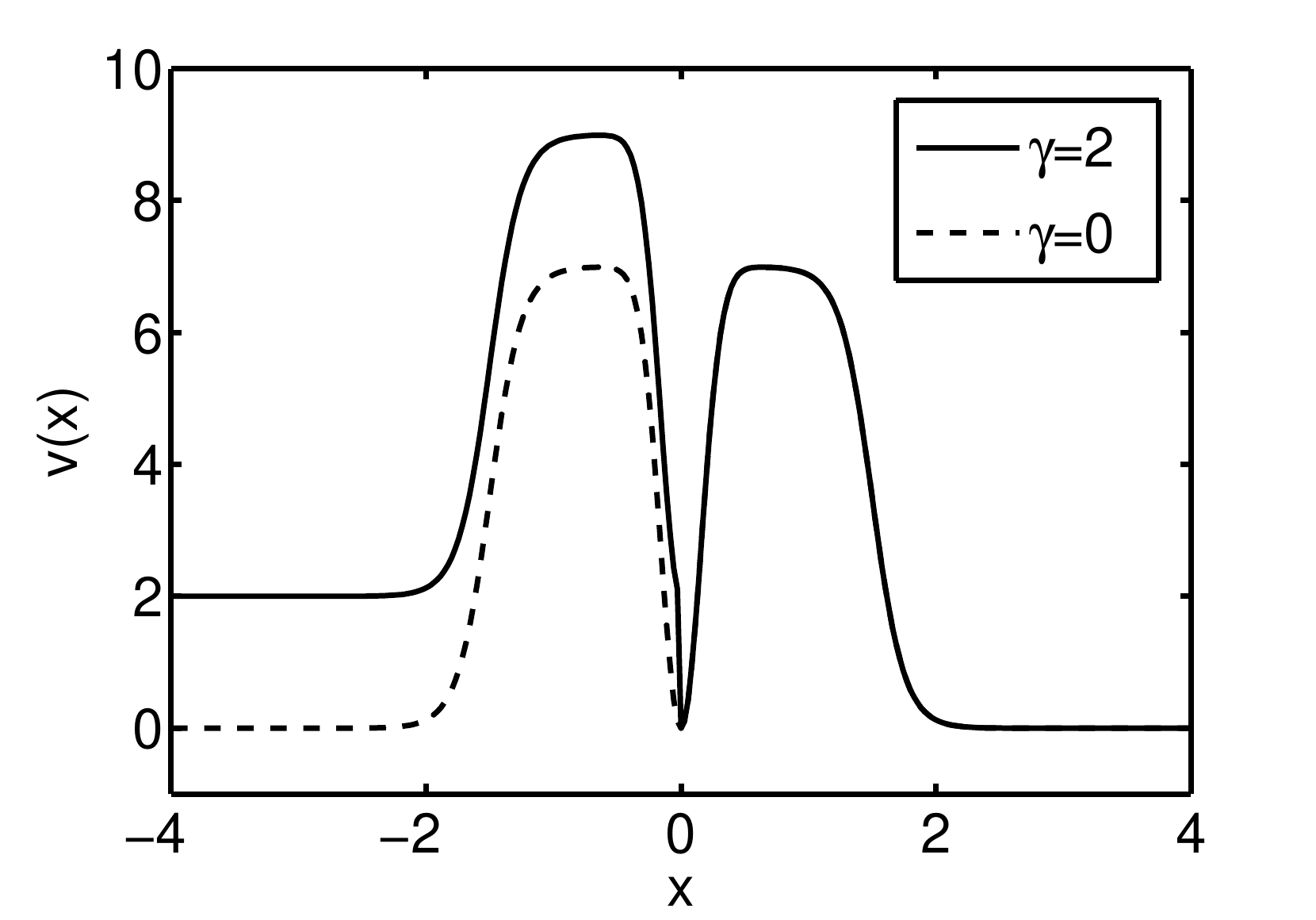}}
\end{center}
\caption{\label{fig:steppot} Model potential to observe the behavior of the complex density in assymetric potentials ($a=7$, $b=0.05$, $c=4$, $d=1.5$).  Note that $\gamma$ is the bias parameter ($\gamma=0$ is unbiased).  }
\end{figure}

\begin{figure}
\begin{center}
\scalebox{0.45}{\includegraphics{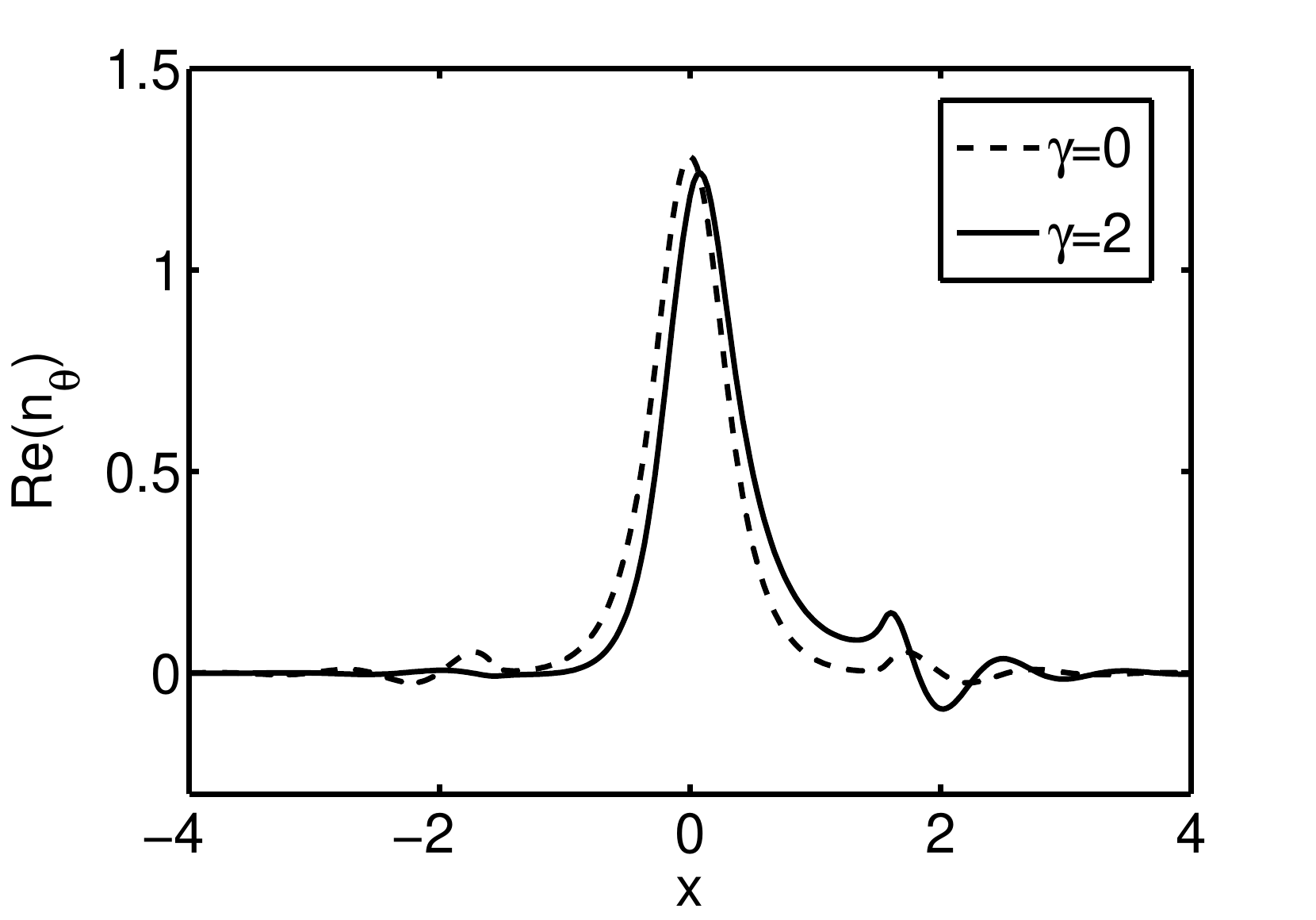}} 
\scalebox{0.45}{\includegraphics{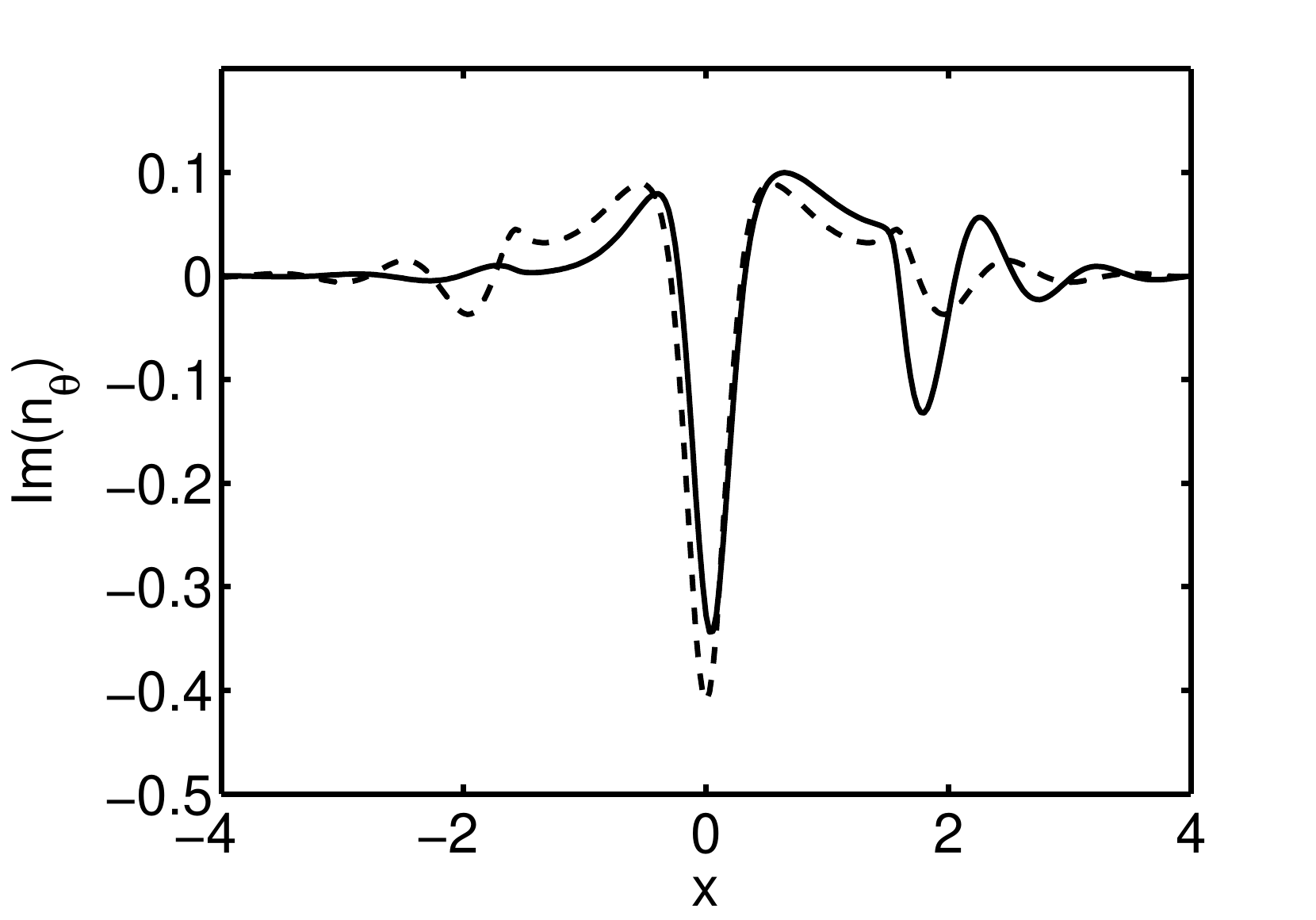}}
\end{center}
\caption{\label{fig:assymden} Real and imaginary one electron complex densities.  One with an assymetric bias and one without ($a=7$, $b=0.05$, $c=4$, $d=1.5$).}
\end{figure}

Along with the magnitude and phase of $n_\theta$, the structure of the complex function itself can provide physical insight.  To illustrate this point consider soft-coulomb interacting particles in a 1D potential, $v(x)+ \gamma\Theta (-x)$, where $\gamma$ is a constant and $v(x)$ is given in Eq.~\ref{eq:modelpot}.  This potential only supports resonances and has a bias for the decay from one side to another depending on the value of $\gamma$ (see Fig.~\ref{fig:steppot}).  
The complex density function was calculated, and an example density is included in Figure~\ref{fig:assymden}. Strong oscillations in the tails of the density indicate the geometrical bias of decay.  In this simple example, an electron prefers to exit the potential to the right.  Therefore, this system will tend to interact, or react, more with another system that is brought in from the right rather than from the left.  This structure motivates the use of the complex density function in studying the reactivity of negative ions.  In three dimensions a gradient of the complex density could be employed to see local areas where there is a geometrical bias of decay.  It is out of these critical areas that the metastable system would tend to donate an electron in a chemical reaction.  Metastable systems are often very reactive due to their diffuse electron clouds, and a systematic theoretical method for exploring reactivity would be of general interest.  
This is related to work from Moiseyev showing how one can extract information about the partial widths from the decay of the complex wavefunctions in assymetric potentials~\cite{PML90, BS82}.  The advantage here is the use of the one-body complex density and not the many-body resonance wavefunctions.  It is possible to develop a useful reactivity theory based on DFRT calculations for the complex density and its various derivatives.

\section{\label{sec:orb} Kohn-Sham Orbital Energies and Lifetimes}
As in DFT, the auxilary KS orbitals or energies in DFRT are assumed to not have a direct physical interpretation.  However, in recent work, the authors have shown that the highest occupied orbital energy and lifetime of DFRT can be related to physical quantities~\cite{WZW11} (similar to Koopmans' theorem of DFT).  Specifically if the system of interest is a $(N+1)$-electron metastable system with a single dominant decay channel, 
\begin{equation}
 (\epsilon_{\sss H}^\theta - \epsilon^{th}) = (-A - \frac{\Gamma}{2}i)~~,
\label{eq:hmt}
\end{equation}
where $\epsilon_{\sss H}^\theta = \varepsilon_{\sss H}-2i\tau_{\sss H}^{-1}$ is the complex Highest Occupied (HOMO) orbital energy of DFRT, $\varepsilon_{\sss H}$ is the HOMO resonance energy or position, $\tau_{\sss H}$ is the HOMO resonance lifetime, $A$ is the negative electron affinity (NEA) of the $N$-electron system, $\Gamma$ is the width of the metastable system, and $\epsilon^{th}$ is the KS ``threshold'' energy defined by,
\begin{equation}
 \epsilon^{th} = E^{th} - \xi [n_\theta] 
\vspace{5pt}
 \label{eq:the}
\end{equation}
where $\xi [n_\theta] = \sum_{i=1}^{N} \epsilon_{i}^\theta +E\Hxc^\theta[n_\theta] -\int d\br v\Hxc^\theta (\br) n_\theta(\br)$.  In other words, the threshold energy in the KS system is the threshold energy in the real system shifted by $-\xi$.  Individually the KS HOMO resonance energy is $\mbox{Re}(\epsilon_{\sss H}^\theta)= \varepsilon_{\sss H} = \mbox{Re}(-A) + \mbox{Re}(\epsilon^{th})$ and the KS HOMO width (inverse lifetime) is $\mbox{Im}(\epsilon_{\sss H}^\theta)= 2\tau_{\sss H}^{-1} = -\frac{\Gamma}{2} + \mbox{Im}(\epsilon^{th})$.  Both the energy and lifetime are real physical quantities shifted by either the real or imaginary part of $\epsilon^{th}$, a factor resulting from Hartree, exchange, and correlation contributions.

For a bound system, the standard Koopmans' theorem of DFT is recovered and the HOMO energy of DFRT is equal to the negative of the ionization potential of the system. 

We emphasize that DFRT is applicable to both bound ground-states and LER's.  Therefore, a system of interest might have some KS orbital energies that are bound and some that are resonances.  For example, consider a metastable negative ion formed by adding an electron to a neutral bound system having a NEA.  A DFRT calculation on the bound parent system would render all orbital energies to have zero imaginary part (i.e. all the Kohn-Sham particles are bound).  A DFRT calculation on the metastable anion would yield all but one KS particle bound.  The remaining particle (the HOMO) would be a resonance with its resonance position and width given by the expressions above.    

In addition to relating the KS HOMO energy to physical quantities, a useful feature of DFT that is preserved in DFRT is the ability to construct the total energy as a corrected sum of orbital energies,
\begin{eqnarray}
E_\theta[n_\theta] &=& \sum_{i=1}^N\left(\varepsilon_i-2i\tau_i^{-1}\right)+E\Hx^\theta[n_\theta] \nonumber \\
 & & -\int d\br v\Hx^\theta (\br) n_\theta(\br)
\label{eq:orbsum}
\end{eqnarray}
This is a useful feature for practical implementations.

A peculiarity of the Kohn-Sham lifetimes can be seen by considering a system of non-interacting particles in a potential that only supports resonances.  In this case, the Kohn-Sham potential of DFRT is just the complex-scaled external potential, and Eq.~\ref{eq:orbsum} simply reduces to the sum over orbital energies.  The lifetime of the full system is ${\cal{L}}^{-1} = \sum_{i=1}^N \tau_i^{-1}$.  For two non-interacting particles this implies that ${\cal{L}}=\tau_{\sss H}/2$ and the lifetime of the system of two particles is less than that of the system with one particle even though the particles are non-interacting.

\section{\label{sec:func} Scaling of Exact and Approximate Functionals}
Any implementation of DFRT must approximate the universal functional of Eq.~\ref{eq:fn}.  Related work by Ernzerhof \cite{E06} and physical intuition suggest that bound ground-state functionals are applicable.  If the energy functional is partitioned as in Eq.~\ref{e:chypothesis}, the complex-coordinate scaling of the kinetic and Hartree functionals follows immediately from the coordinate scaling of the kinetic and interaction operators: $T_s^\theta[n_\theta]=e^{-2i\theta} T_s[n_\theta]$  and $E\H^\theta[n_\theta]=e^{-i\theta}E\H[n_\theta]$, where $T_s[n_\theta]$ and $E\H[n_\theta]$ are the standard non-interacting kinetic energy and Hartree DFT functionals evaluated at the complex densities.  The scaling of the exchange and correlation functional is not immediately apparent. 

To study the complex-scaling of functionals, one can take advantage of the fact that DFRT can be applied to bound ground-states, not just LER's.  For a system with a bound ground state, various coordinate scaling relations ($\br \rightarrow \alpha \br$ for $\alpha > 0$) are known for the functionals of ground-state DFT~\cite{gfg}.  Our strategy to derive exact relations between DFT and DFRT functionals, is to consider the complex-coordinate-scaling transformation, $\br \rightarrow \br e^{i \theta}$, keeping in mind that DFT and DFRT give the exact same energy for a system with a bound ground-state.  For such a system, this correspondence between the energies of DFT and DFRT implies:  
\begin{equation}
 F[n] + \int d\br \ v(\br) n(\br) = F_\theta [n_\theta] + \int d\br \ v(\br e^{i \theta}) n_\theta(\br)
\label{eq:same}
\end{equation}
where $F[n]$ and $F_\theta[n_\theta]$ are the universal functionals of DFT and DFRT respectively, and $n(\br)$ and $n_\theta(\br)$ are the real-valued density of DFT and the complex-valued density of DFRT corresponding to the ground-state.  Once the complex-scaling relations of the functionals are in place, the assumption is made that the scaling does not change for a LER, even though Eq.~\ref{eq:same} is no longer valid.  This assumption is reasonable because the functional (Eq.~\ref{eq:fn}) of DFRT is applicable to both bound and LER states.

Now, consider the change in the ground-state functionals with a complex-scaling of the wavefunction and density coordinates.  As in standard DFT, it is required that the scaled density integrates to the number of electrons. 
\begin{equation}
 \int d\br \ n_\theta(\br) = \int d\br \ n (\br) = N
\label{eq:scalednorm}
\end{equation}
In three dimensions this means:
\begin{equation}
  n(\br) \rightarrow n_\theta(\br) \equiv e^{3 i \theta} n(\br e^{i \theta})
 \label{eq:denscaled}
\end{equation}
so that the transformation (coordinate scaling) preserves the number of electrons.  For the non-interacting kinetic energy, $T_s^\theta [n_\theta]$, the ground-state wavefunction that yields $n_\theta$ according to Eq.~\ref{eq:scalednorm} is given by
\begin{equation}
  \Psi_{\theta} (\br_1,\ldots,\br_N) \equiv e^{3 i N \theta / 2} \Psi (\br_1 e^{i \theta},\ldots,\br_N e^{i \theta})
 \label{eq:psiscale}
\end{equation}
Then,
\begin{equation}
  ( \Psi_{\theta} | \hat{T} | \Psi_{\theta} ) = e^{2 i \theta} \langle \Psi | \hat{T} | \Psi \rangle
\label{eq:tscale}
\end{equation}
where the ``c-product,'' indicated by $( \Psi_{\theta} | \Psi_{\theta} )$, is used to emphasize that the imaginary part of $\Psi_{\theta}$ that becomes complex purely from the transformation is not conjugated~\cite{M11}.  Eq.~\ref{eq:tscale} shows that the functional has a coefficient of $e^{2 i \theta}$ from the scaling of the wavefunction. Yet, the complex-coordinate scaling leaves bound state energies untouched.  Therefore, the scaling factor in the complex-scaled functional of DFRT, which is also applicable to bound states, must compensate for this factor of $e^{2 i \theta}$,
\begin{equation}
 T_s^\theta[n_\theta]=e^{-2i\theta} T_s[n_\theta]
\end{equation}
Next, consider the Hartree functional.  Like the kinetic functional, the relation of the complex Hartree functional of DFRT, $E\H^\theta[n_\theta]$, to the standard ground-state Hartree functional, $E\H[n]$, follows from the coordinate scaling of the interaction operator.  To emphasize this point, the change in $E\H[n]$ caused by a scaling of the density coordinates is considered.  By directly substituting $n_\theta(\br)$, defined in Eq.~\ref{eq:denscaled}, into $E\H[n]$,
\begin{equation}
 E\H[n_\theta] = e^{i \theta} E\H[n]
\label{eq:horgin}
\end{equation}
and the functional $E\H^\theta[n_\theta]$ must compensate for this factor,
\begin{equation}
 E\H^\theta[n_\theta]=e^{-i\theta}E\H[n_\theta]
 \label{eq:hscale}
\end{equation}
The exchange functional, $E\x[n]$, is defined by,
\begin{equation}
 E\x[n] = \langle \Psi^{s} | \hat{V}_{ee} | \Psi^{s} \rangle - E\H[n]
\end{equation}
where $\Psi^{s}$ is the Kohn-Sham wavefunction that yields $n(\br)$ and minimizes the expectation value of $\hat{T}$.  Using Eq.~\ref{eq:horgin} together with Eq.~\ref{eq:psiscale},
\begin{equation}
 E\x[n_\theta] = e^{i \theta} E\x[n] 
\end{equation}
and,
\begin{equation}
 E\x^\theta[n_\theta] = e^{-i\theta}E\x[n_\theta]
 \label{eq:xscale}
\end{equation}
Lastly, consider the correlation functional, $E\c[n]$, defined by,
\begin{equation}
 E\c[n] = \langle \Psi | \hat{T} + \hat{V}_{ee} | \Psi \rangle - \langle \Psi^s | \hat{T} + \hat{V}_{ee} | \Psi^s \rangle
\end{equation}
where $\Psi$ yields $n$ and minimizes the expectation value of $(\hat{T} + \hat{V}_{ee})$.  The scaling of this functional is more complicated.  Write
\begin{equation}
  \Psi_{\theta} (\br_1,\ldots,\br_N) = e^{3 i N \theta / 2} \Phi (\br_1 e^{i \theta},\ldots,\br_N e^{i \theta})
 \label{eq:psi2scale}
\end{equation}
where $\Phi (\br_1,\ldots,\br_N )$ is unknown but yields $n(\br)$.  Using what we have already learned above about the scaling of the kinetic and Hartree functionals,
\begin{equation}
( \Psi_\theta | \hat{T} + \hat{V}_{ee} | \Psi_\theta ) = e^{2 i \theta} ( \Phi | \hat{T} + e^{-i \theta} \hat{V}_{ee} | \Phi )
\end{equation}
If $\Phi$ is then chosen to minimize the real and imaginary parts of the bi-expectation value of $\hat{T} + e^{-i \theta} \hat{V}_{ee}$,
\begin{eqnarray}
E\c[n_\theta(\br)] &=& ( \Psi_\theta | \hat{T} + \hat{V}_{ee} | \Psi_\theta ) - ( \Psi_{\theta}^s | \hat{T} + \hat{V}_{ee} | \Psi_{\theta}^s ) \nonumber \\
 &=& e^{2 i \theta} ( \Phi | \hat{T} + e^{-i \theta} \hat{V}_{ee} | \Phi )  \nonumber \\
 & & - e^{2 i \theta} \langle \Psi_0 | \hat{T} + e^{-i \theta} \hat{V}_{ee} | \Psi_0 \rangle \nonumber \\
 &=& e^{2 i \theta} E_{{\sss C},\lambda} [n]
\label{eq:coupcon}
\end{eqnarray}
where $\lambda = e^{-i \theta}$, and $E_{{\sss C},\lambda}$ is the correlation energy functional for a coupling-constant Hamiltonian:
\begin{equation}
 \hat{H}(\lambda) = \hat{T} + \lambda \hat{V}_{ee} + \int d\br \ v(\br,\lambda) \hat{n}(\br) 
\end{equation}
and $v(\br,\lambda)$ is chosen such that the density of the ground-state is exactly $n(\br)$ independent of $\lambda$.  The complex correlation function of DFRT would then obey
\begin{equation}
 E\c^\theta[n_\theta] = e^{-2 i \theta} E_{{\sss C},\lambda} [n_\theta]
\end{equation}
Note, it is not proven that the potential $v(\br,\lambda)$, and hence the coupling-constant system, exists for this complex choice of $\lambda$.  This system is quite different from the coupling-constant system typically used to derive a scaling relationship for the ground-state DFT correlation functional.  One can see that $\hat{H}(\lambda)$ is non-Hermitian and therefore the bi-expectation value is used in Eq.~\ref{eq:coupcon}.  $v(\br,\lambda)$ must be chosen such that the complex-valued function $\Phi$ minimizes the real and imaginary parts of the bi-expectation value of $\hat{T} + e^{-i \theta} \hat{V}_{ee}$ and yields $n(\br)$, a real density, from the bi-expectation value $(\Phi | \hat{n} (\br) | \Phi )$. Assuming the existence of such a system shows a possible mapping between the complex correlation functional of DFRT and the correlation functional of ground-state DFT.  Therefore, the natural choice of ground-state functionals evaluated at the complex density for DFRT seems reasonable provided the correct scaling factors are included.

Even though the use of ground-state functionals in a DFRT implementation seems promising, the analytic form of some approximate functionals could create issues.  For example, consider the simple LDA appoximation to the exchange energy,
\begin{equation}
 E\x^{{\sss LDA}}[n_\theta] = C\x \int d\br \ n_\theta^{4/3} (\br)
\end{equation}
where $C\x = -(3/4)(2/\pi)^{1/3}$.  The $(4/3)$ power in the functional is not uniquely defined for complex functions.  This problem was considered recently by Ernzerhof \emph{et al.} who concluded that the non-uniqueness is removed if the complex density behaves continuously as the imaginary part of the potential is reduced to zero~\cite{ZE11}.  One should be careful to analyze the form of approximate ground-state functionals before their implementation in an approximate DFRT.

Along with possible uniqueness issues, any approximate functional used in an implementation of DFRT could render a bound system unbound or an unbound system bound.  In other words, errors in the approximate DFRT functionals could give the energy of a bound state a non-zero imaginary part, hence indicating that the state has a finite lifetime.  In this case, the imaginary part of the energy measures the error induced by an approximate complex functional.  These types of errors are expected to be most visible when the ground-state of a system is weakly bound or for a metastable system that has a very long lifetime.

\section{Conclusion}

In this work, the fundamental quantities of DFRT and practical considerations of its implementation have been explored.  Results of model calculations show that sensitivity to the complex-scaling parameter can be avoided via the use of a non-interacting auxilary system, the Kohn-Sham system.  These one-particle equations can be liberated of any $\theta$-dependence by using a large, but still reasonable, basis set or fine grid.  The stucture and significance of the orbital energies and lifetimes of this Kohn-Sham system have been studied for some relevant cases. Some systems, such as the LER of a system with a NEA, might have a mix of both bound and metastable Kohn-Sham particles.  The HOMO energy is related to the NEA, the width of the LER, and the threshold energy of the system.  In addition, physical interpretation has been assigned to features of complex density functions such as oscillations which indicate geometrical bias of decay, and a reactivity theory based on these functions has been motivated via model system calculations.  Lastly, the complex-scaling of ground-state DFT functionals has been studied, a neccessary step towards general use of DFRT.  Scaling relationships have been derived which show how the DFRT functionals can be written as the DFT functional evaluated at the complex density multiplied by a complex constant.  An implementation of DFRT for real systems of interest and a time-dependent extension are forthcoming.

\begin{acknowledgments}
This work was supported by the Purdue Research Foundation, the Donors of the American Chemical Society Petroleum Research Fund grant No.PRF\# 49599-DNI6, and the National Science Foundation CAREER program grant No.CHE-1149968.
\end{acknowledgments}

%

\end{document}
%